\documentclass[epj]{svjour}
\usepackage{latexsym}
\usepackage{graphics}
\usepackage{amsmath}
\usepackage{amssymb}
\usepackage{indentfirst}
\usepackage[all]{xy}
\usepackage{multirow}
\usepackage{booktabs}
\usepackage{makecell}
\begin{document}
%
\title{Theoretical Investigation of the Black-body Zeeman Shift for Microwave Atomic Clocks}
%
\author{Jize Han\inst{1}    \and Yani Zuo\inst{1}   \and Jianwei Zhang\inst{2, 3, a}    \and Lijun Wang\inst{1, 2, 3}
%
}
    \titlerunning{ Theoretical Investigation of the Black-body Zeeman Shift for Microwave Atomic Clocks}
    \authorrunning{Jize Han et al.}
\mail{\emph{zhangjw@tsinghua.edu.cn}}

%
%
\institute{Department of Physics, Tsinghua University, Beijing 100084 \and State Key Laboratory of Precision Measurement Technology and Instruments, Tsinghua University, Beijing 100084 \and Department of Precision Instruments, Tsinghua University, Beijing 100084}
%
\date{Received: date / Revised version: date}
%
\abstract{
       With the development of microwave atomic clocks, the Zeeman shifts for the spectral lines of black-body radiation need to be investigated carefully. In this Letter, the frequency shifts of hyperfine splittings of atomic ground states due to the magnetic field of black-body radiation are reported. The relative frequency shifts of different alkali atoms and alkali-like ions, which could be candidates of microwave atomic clocks, were calculated. The results vary from $-0.977\times10^{-17}[T(K)/300]^{2}$ to $-1.947\times10^{-17}[T(K)/300]^{2}$ for different atoms considered. These results are consistent with previous work but with greater precision, detailed derivations, and a clear physical picture.
    \PACS{
      {}{No more than four PACS codes should be provided} 
     }
}
%
\maketitle
\section{Introduction}
\label{intro}
    For any atomic system, the energy levels are shifted by black-body radiation (BBR). These shifts may be ignored because their values are small. With the development of atomic clocks, they become one of the main components of frequency shifts, and indubitably have to be considered. Regarding microwave and optical atomic clocks, there are many theoretical $^{[1-8]}$ and experimental $^{[9,10]}$ studies of BBR shifts. However, most of those studies focus on BBR Stark (BBRS) shifts, because they are much larger than the BBR Zeeman (BBRZ) shifts pertinent for microwave atomic clocks. To the best of our knowledge, there are only a few theoretical studies$^{[11-12]}$ on the BBRZ shifts and no experimental results. To date, the frequency uncertainty of the state-of-the-art cesium fountain has been decreased to order 1E-16. It is worthwhile considering more carefully the BBRZ shifts for microwave atomic clocks as the uncertainties in frequency may improved to order 1E-17 in the next decade.
    \par
    In this Letter, we give a detailed derivation of the BBRZ shifts for the ground state of alkali atoms and alkali-like ions, which could be used in microwave atomic clocks. This theoretical investigation of the BBRZ shifts may be important for the development of microwave atomic clocks.
\section{Theory}
\label{sec:2}
\subsection{Black-body radiation field}
\label{sec:2.1}
    \par
    According to Planck's radiation law,
\begin{equation}
\begin{aligned}
    \int_{0}^{+\infty}u(\omega)d\omega
    &=\int_{0}^{+\infty}\frac{\hbar}{\pi^2c^3}
       \frac{\omega^3}{e^{\frac{\hbar\omega}{kT}}-1}dw\\
    &=\frac{1}{2}\varepsilon_0\langle E^2(t)\rangle+\frac{1}{2\mu_0}\langle B^2(t)\rangle,
\end{aligned}
\end{equation}
where $u(\omega)d\omega$ is the field energy density in the bandwidth $d\omega$ around angular frequency $\omega$, $\langle E^2(t)\rangle$ and $\langle B^2(t)\rangle$ are the mean-squares of the electric and magnetic fields, respectively. The mean-squares of the fields are defined as
\begin{equation}
    \langle B^2(t)\rangle=\frac{1}{2}\int_{0}^{+\infty}B^2(\omega)d\omega
\end{equation}
    and
\begin{equation}
    \langle E^2(t)\rangle=\frac{1}{2}\int_{0}^{+\infty}E^2(\omega)d\omega.
\end{equation}
    Here, $E^2(\omega)$ and $B^2(\omega)$ are the field energy densities in a bandwidth $d\omega$ around $\omega$ of the electric and magnetic fields, respectively.
    \par
    According to the theory of electromagnetism, the fields in an electromagnetic wave have the same energy. Thus, $B^2(\omega)d\omega$ is
\begin{equation}
    B^2(\omega)d\omega=\frac{2\mu_0\hbar}{\pi^2c^3}
       \frac{\omega^3}{e^{\frac{\hbar\omega}{kT}}-1}dw.
\end{equation}
\subsection{Radiation field interacting with two-level atoms}
\label{sec:2.2}
    \par
     We consider a simple situation where a magnetic field oscillates at angular frequency $\omega$ interacting with two-level atoms. The evolution of the system is governed by the time-dependent Schr\"odinger equation
\begin{equation}
    i\hbar\frac{d\Psi}{dt}=\widehat{H}\Psi,
\end{equation}
    with Hamiltonian written as
\begin{equation}
    \widehat{H}=\widehat{H}_A+\widehat{H}_{AF}(t),
\end{equation}
     where
\begin{equation}
    \widehat{H}_{AF}(t)= \boldsymbol{\mu}\cdot\boldsymbol{B}cos\omega{t}.
\end{equation}
    Here, $\widehat{H}_A$ is the Hamiltonian of the free atom, $\widehat{H}_{AF}$ the atom--field interaction Hamiltonian, $\boldsymbol{\mu}$ the magnetic moment, and $\boldsymbol{B}$ the amplitude of the monochromatic field.
    \par
    The wave function associated with $\widehat{H}$ is a superposition of the two eigenfunctions,
\begin{equation}
    \Psi(r,t)=C_g(t)\Psi_g(r)e^{-i\omega_gt}+C_e(t)\Psi_e(r)e^{-i\omega_et},
\end{equation}
where $\omega_g$ and $\omega_e$ are the frequencies of the ground and excited state, and $C_g$ and $C_e$ the coefficients of overlap for the two states, respectively.
    \par
    We may next derive two coupled equations for these coefficients,
\begin{equation}
\left\{
	\begin{aligned}
    	i\dot{C}_g&=C_e[e^{i(\omega-\omega_0)t}+e^{-i(\omega+\omega_0)t}]\frac{\Omega}{2}\\
	i\dot{C}_e&=C_g[e^{-i(\omega-\omega_0)t}+e^{i(\omega+\omega_0)t}]\frac{\Omega}{2},
	\end{aligned}
\right.
\end{equation}
    where $\Omega=\langle g|\boldsymbol{\mu}\cdot\boldsymbol{B}|e\rangle/{\hbar}$ is the magnetic Rabi frequency, and $\omega_0=\omega_e-\omega_g$ is the resonance frequency between the two levels.
    \par
    If the frequency of the radiation field $\omega$ is near resonance, $|\omega-\omega_0|\ll\omega_0$, we use the rotating-wave approximation (RWA) to simplify Eq.~(9), and we have
\begin{equation}
\left\{
\begin{aligned}
    i\dot{C}_g &= C_ee^{i(\omega-\omega_0)t}\frac{\Omega}{2}\\
    i\dot{C}_e &= C_ge^{-i(\omega-\omega_0)t}\frac{\Omega}{2}.
\end{aligned}
\right.
\end{equation}
    \par
    After solving the secular equations, Eq.~(10), the frequency shifts induced by the rotating term is$^{[13]}$
\begin{equation}
\begin{aligned}
    \Delta\omega_{rot}&=-\frac{(b_e^2+b_g^2)\mu_B^2B^2}{4(\omega-\omega_0)}\\
    &=-\frac{b^2\mu_B^2B^2}{4(\omega-\omega_0)},
\end{aligned}
\end{equation}
    where $b=|\langle g|\boldsymbol{\mu}|e\rangle| \slash \mu_B$ is the coupling coefficient, $b_g$ and $b_e$ are the coupling coefficients of the ground and excited states, respectively, and the total coupling coefficient $b=\sqrt{b_e^2-(-b_g^2)}$.
    \par
    For far detuning, $|\omega-\omega_0|\approx\omega_0$ or $\omega$, applying the RWA is not valid. Instead, the counter-rotating terms are given by
\begin{equation}
\left\{
\begin{aligned}
    i\dot{C}_g &= C_ee^{-i(\omega+\omega_0)t}\frac{\Omega}{2}\\
    i\dot{C}_e &= C_ge^{i(\omega+\omega_0)t}\frac{\Omega}{2}.
\end{aligned}
\right.
\end{equation}
    \par
    We may treat the counter-rotating terms as another monochromatic field with frequency $-\omega$. The frequency shift has the same form except $\omega$ is replaced by $-\omega$. Thus, the frequency shift caused by the counter-rotating term is $^{[14]}$
\begin{equation}
    \Delta\omega_{crot}=\frac{b^2\mu_B^2B^2}{4(\omega+\omega_0)\hbar^2}.
\end{equation}
    \par
    Combining Eqs.~(11) and (13), the total frequency shift of the far-detuning situation are
\begin{equation}
\begin{aligned}
    \Delta\omega_{fdet}&=\Delta\omega_{rot}+\Delta\omega_{crot}\\
    &=-\frac{\omega_0b^2\mu_B^2B^2}{2(\omega^2-\omega_0^2)\hbar^2}.
\end{aligned}
\end{equation}
    \par
    For near resonance, there is a more accurate expression for the frequency shift$^{[15,16]}$. Therefore, the complete expression of the ground state frequency shift for the two distinct situations is
\begin{equation}
\Delta\omega=
\left\{
\begin{aligned}
    &-\frac{\omega_0b^2\mu_B^2B^2(\omega)}{2(\omega^2-\omega_0^2)\hbar^2}\\
    &\omega \in [0, \omega_0-\delta\omega_0] \cup [\omega_0+\delta\omega_0,\infty]\\
    \\
    &-\frac{b^2\mu_B^2B^2(\omega)}{2\hbar^2}\cdot\frac{\omega-\omega_0}{(\omega-\omega_0)^2+\frac{b^2\mu_B^2B^2(\omega)}{\hbar^2}},\\
    &\omega \in [\omega_0-\delta\omega_0, \omega_0+\delta\omega_0]
\end{aligned}
\right.
\end{equation}
    with the excited state having the opposite sign. In Eq.~(15), we choose $\omega_0\mp\delta\omega_0$ as the lower and upper limits of the near-resonance situation. Without loss of generality, we set $\delta\omega_0=0.01\omega_0$ for the calculation.
\subsection{Calculation of the coupling coefficients}
\label{sec:2.3} 
    \par
    To calculate $\Delta\omega$, we need to calculate the coupling coefficient,
\begin{equation}
    b=\frac{|\langle g|\boldsymbol{\mu}|e\rangle|}{\mu_B}
=\frac{|\langle \boldsymbol{F}m_F|\boldsymbol{\mu}|\boldsymbol{F^{'}}m_F^{'}\rangle|}{\mu_B},
\end{equation}
    between the ground state and the excited state. Because the total angular momentum $\boldsymbol{F}$ and its magnetic quantum number $m_F$ are different for $g$ and $e$, we cannot obtain the coupling coefficient $b$ directly using the coupling formula for the total electronic angular momentum g-factor $g_J$, and nuclear spin angular momentum g-factor $g_I$. Instead, we need to transform the coupled representation into an uncoupled representation using the Clebsch--Gordan coefficients,
\begin{equation}
\begin{aligned}
    |\boldsymbol{J}\boldsymbol{I};\boldsymbol{F}m_F\rangle=&\\
    \sum_{m_J=-J}^{J}\sum_{m_I=-I}^{I}&|\boldsymbol{J}m_J;\boldsymbol{I}m_I\rangle\langle \boldsymbol{J}m_J;\boldsymbol{I}m_I|\boldsymbol{J}\boldsymbol{I};\boldsymbol{F}m_F\rangle.
\end{aligned}
\end{equation}
    \par
		For example, the clock transition for atoms with nuclear spin $\boldsymbol{I}=1\slash2$ is the hyperfine splittings $^2S_{1\slash2}$ $\boldsymbol{F}=0$, $m_F=0$ to $^2S_{1\slash2}$ $\boldsymbol{F}=1$, $m_F=0$; the energy levels are shown in Fig.~1.\\
\begin{figure}
\centering
\resizebox{0.45\textwidth}{!}{%
    \includegraphics{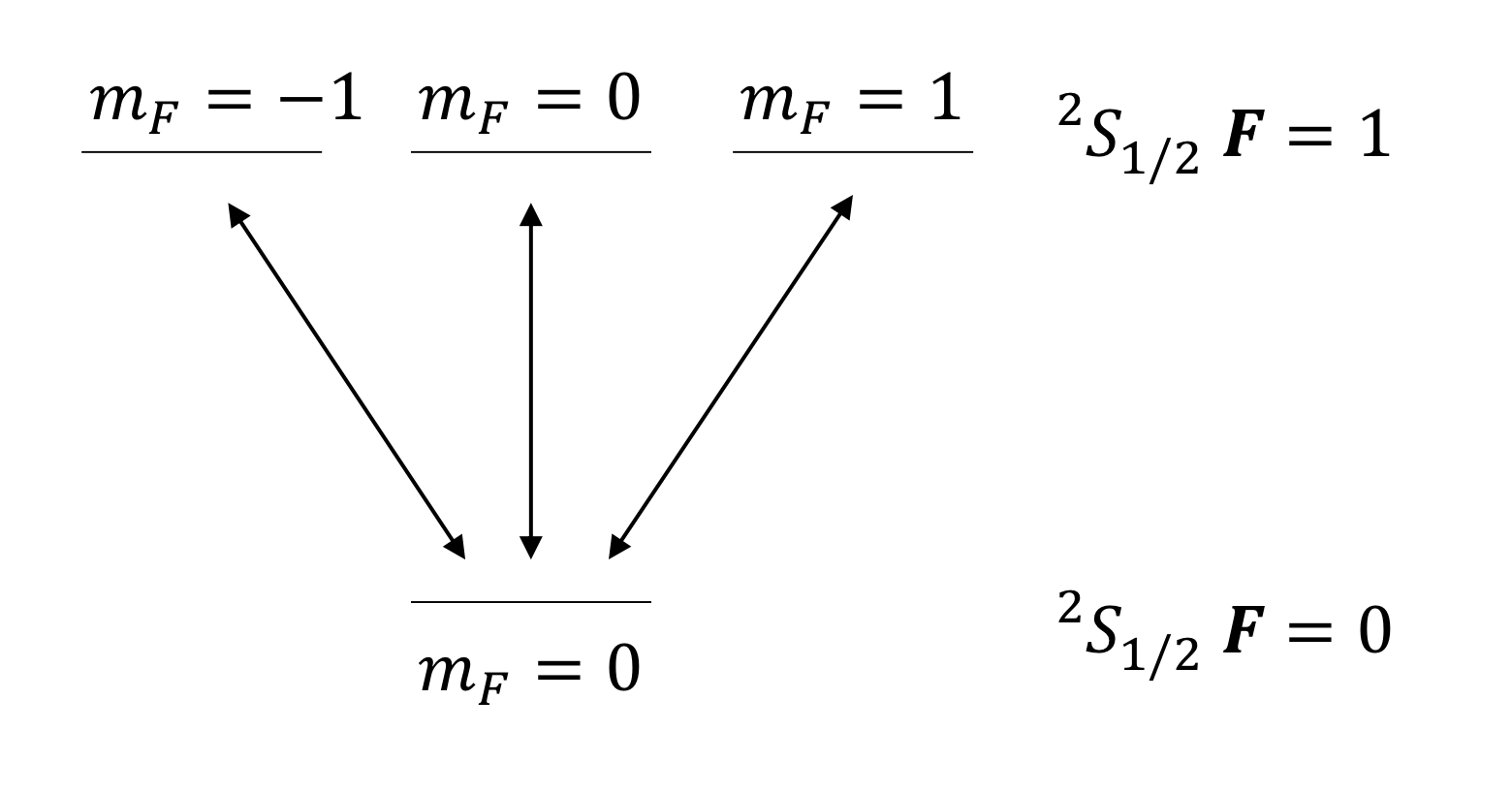}
}
\caption{Energy levels for the ground state of $\boldsymbol{I}=1\slash2$ atom}
\label{fig:1}       
\end{figure}
    \par
    As shown in Fig.~1, we need to consider all other states that could influence our target states. In this instance, there are three transitions that shift the ground state level of the clock transition: $|00|\rightarrow|11\rangle$, $|00\rangle\rightarrow|10\rangle$, and $|00\rangle\rightarrow|1-1\rangle$.
    \par
    For the energy transition $|00\rangle\rightarrow|11\rangle$, the uncoupled representations of each levels are shown in Fig.~2.\\
\begin{figure}
\centering
\resizebox{0.45\textwidth}{!}{%
    \includegraphics{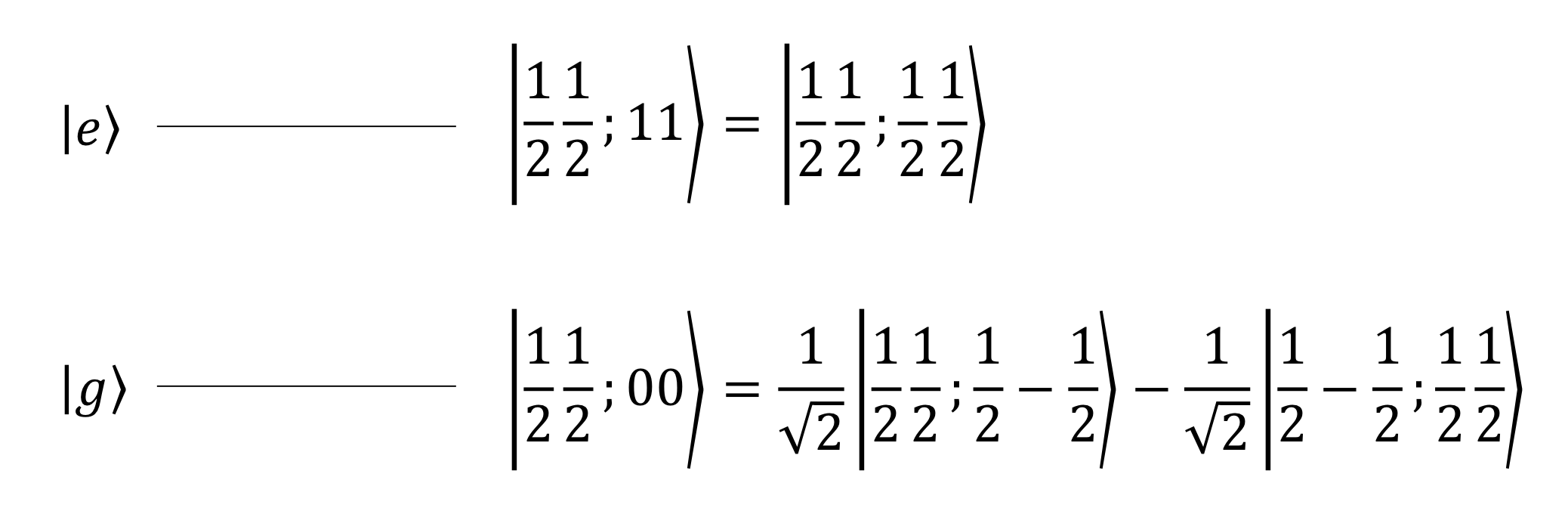}
}
\caption{Uncoupled representations of $|00\rangle$ $|11\rangle$}
\label{fig:1}       
\end{figure}
    \par
    Thus, the coupling coefficient $b_g$ is
\begin{equation}
\begin{aligned}
    b_g &=\langle00|g_J\boldsymbol{J}+g_I\boldsymbol{I}|11\rangle\\
        &=\bigg(\frac{1}{\sqrt{2}}\bigg\langle\frac{1}{2}\frac{1}{2};\frac{1}{2}-\frac{1}{2}\bigg|-\frac{1}{\sqrt{2}}\bigg\langle\frac{1}{2}-\frac{1}{2};\frac{1}{2}\frac{1}{2}\bigg|\bigg)\\
        &\quad\cdot(g_J\boldsymbol{J}+g_I\boldsymbol{I})\bigg(\bigg|\frac{1}{2}\frac{1}{2};\frac{1}{2}\frac{1}{2}\bigg\rangle\bigg)\\
        &=\frac{1}{2\sqrt{2}}(g_J-g_I).
\end{aligned}
\end{equation}
    \par
    Using the same methods, we determined the coupling coefficients for $|00\rangle\rightarrow|10\rangle$ and $|00\rangle\rightarrow|1-1\rangle$. The results of $b_g$ are ${(g_J-g_I)}\slash{2}$ and ${(g_J-g_I)}\slash{2\sqrt{2}}$, respectively.
    \par
    Only the $|10\rangle\rightarrow|00\rangle$ transition shifts the excited state of the clock transition; specifically, we find $b_e$ is ${(g_J-g_I)}\slash{2}$.
    \par
    The total coupling coefficient of the clock transition with $\boldsymbol{I}=1\slash2$ is
\begin{equation}
    b=\sqrt{b_e^2-(-b_g^2)}={\frac{\sqrt{3}}{2}(g_J-g_I)}.
\end{equation}
    \par
    For different nuclear spins, the coupling coefficients are calculated in the same way; the results are listed in Table 1.
    \\
\begin{table}
\caption{Coupling coefficients for different nuclear spins}
\label{tab:1}       
\begin{tabular}{cll}
\hline\noalign{\smallskip}
    \makecell[c]{Nuclear \\spin }
            & \makecell[c]{Alkali atoms and\\Alkali-like ions}
                     & \makecell[l]{Total Coupling \\Coefficient $b$}\\
\noalign{\smallskip}\hline\noalign{\smallskip}$\frac{1}{2}$ &
\makecell[l]{$^1$H, $^3$H, $^3$He$^+$, $^{111}$Cd$^+$ \\
$^{113}$Cd$^+$, $^{171}$Yb$^+$, $^{199}$Hg$^+$}&
\makecell[l]{$\frac{\sqrt{3}}{2}(g_J-g_I)$}\\
\noalign{\smallskip}\hline\noalign{\smallskip}$\frac{3}{2}$ &
\makecell[l]{$^7$Li, $^9$Be$^+$, $^{39}$K, $^{41}$K, $^{201}$Hg$^+$ \\
$^{23}$Na,  $^{87}$Rb, $^{135}$Ba$^+$, $^{137}$Ba$^+$}&
\makecell[l]{$\frac{1}{\sqrt{3}}\{[(3+\sqrt{2})g_J^2$ \\
$ -\frac{7g_Jg_I}{6}+\frac{3g_I^2}{4}]\}^{\frac{1}{2}}$}\\
\noalign{\smallskip}\hline\noalign{\smallskip}$\frac{5}{2}$ &
$^{25}$Mg$^+$, $^{67}$Zn$^+$, $^{85}$Rb, $^{173}$Yb$^+$ &
\makecell[l]{$\frac{1}{\sqrt{3}}\{[(3+\sqrt{2})g_J^2$ \\
$-\frac{7g_Jg_I}{6}+\frac{3g_I^2}{4}]\}^{\frac{1}{2}}$}\\
\noalign{\smallskip}\hline\noalign{\smallskip}$\frac{7}{2}$ &
$^{43}$Ca$^+$, $^{133}$Cs & \makecell[l]{$\frac{1}{2\sqrt{2}}\{[(8+\sqrt{15})g_J^2$\\
$-9g_Jg_I+6g_I^2]\}^{\frac{1}{2}}$}\\
\noalign{\smallskip}\hline\noalign{\smallskip}$\frac{9}{2}$ &
$^{87}$Sr$^+$& \makecell[l]{$\frac{1}{2\sqrt{5}}\{[(5+\sqrt{6})g_J^2$ \\
$-22g_Jg_I+15g_I^2]\}^{\frac{1}{2}}$}\\
\noalign{\smallskip}\hline
\end{tabular}
\end{table}
\section{Results and discussion}
\label{sec:3}
    \par
    The BBR field is a broad-band radiation field with a uniform spatial distribution. There is only ${1}\slash{3}$ of total radiation field parallel to the direction of the static magnetic field, which induces the $m_F=0\rightarrow m_F=0$ transition. The remaining part of the radiation field is orthogonal to the static magnetic field, and may decompose into left and right circular-polarized light, which induces the $m_F=0\rightarrow m_F=\pm1$ transitions$^{[17]}$. To obtain the total BBRZ shifts, Eq.~(15) is integrated for all $\omega$, and multiplied by ${1}\slash{3}$ to account for the spatial distribution,
\begin{equation}
\begin{aligned}
    \frac{\Delta\omega}{\omega_0}=&-\int_0^{\omega_0-\delta\omega_0}\frac{b^2\mu_B^2B^2(\omega)}{6\hbar^2(\omega^2-\omega_0^2)}dw\\
        &-\int_{\omega_0-\delta\omega_0}^{\omega_0+\delta\omega_0}\frac{b^2\mu_B^2B^2(\omega)}{6\omega_0\hbar^2}\cdot\frac{\omega-\omega_0}{(\omega-\omega_0)^2+\frac{b^2\mu_B^2B^2(\omega)}{\hbar^2}}dw\\
        &-\int_{\omega_0+\delta\omega_0}^{+\infty}\frac{b^2\mu_B^2B^2(\omega)}{6\hbar^2(\omega^2-\omega_0^2)}dw.
\end{aligned}
\end{equation}
    \par
    Using Eq.~(20), we then calculated the frequency shifts for each candidate atom of a microwave atomic clock. The data required concerning the atomic structure are cited from Refs.~[18--21]. The results are listed in Table 2.
\begin{table}
\caption{Relative BBRZ shifts for the ground states of alkali atoms and alkali-like ions}
\label{tab:2}       
\begin{tabular}{ccc}
\hline
    \makecell[c]{Nuclear \\spin $\boldsymbol{I}$} & \makecell[c]{Alkali atoms and \\Alkali-like ions} & \makecell[c]{Relative BBRZ Frequency \\shifts $\frac{\Delta\omega}{\omega_0}[\frac{T(K)}{300}]^2 (\times10^{-17})$}\\
\hline\noalign{\smallskip}
    \multirow{7}{*}{${\frac{1}{2}}$}   & $^1$H  & $-0.977$\\
    \noalign{\smallskip}\cline{2-3}\noalign{\smallskip}
                    & $^3$H                   & $-0.977$\\
    \noalign{\smallskip}\cline{2-3}\noalign{\smallskip}
                    & $^3$He$^+$                & $-0.982$\\
    \noalign{\smallskip}\cline{2-3}\noalign{\smallskip}
                    & $^{111}$Cd$^+$            & $-0.981$\\
    \noalign{\smallskip}\cline{2-3}\noalign{\smallskip}
                    & $^{113}$Cd$^+$            & $-0.981$\\
    \noalign{\smallskip}\cline{2-3}\noalign{\smallskip}
                    & $^{171}$Yb$^+$           & $-0.980$\\
    \noalign{\smallskip}\cline{2-3}\noalign{\smallskip}
                    & $^{199}$Hg$^+$            & $-0.979$\\
\noalign{\smallskip}\hline\noalign{\smallskip}
    \multirow{9}{*}{$\frac{3}{2}$}   & $^7$Li & $-1.872$\\
    \noalign{\smallskip}\cline{2-3}\noalign{\smallskip}
                    & $^{23}$Na               & $-1.872$\\
    \noalign{\smallskip}\cline{2-3}\noalign{\smallskip}
                    & $^{39}$K                & $-1.873$\\
    \noalign{\smallskip}\cline{2-3}\noalign{\smallskip}
                    & $^{41}$K                & $-1.873$\\
    \noalign{\smallskip}\cline{2-3}\noalign{\smallskip}
                    & $^{87}$Rb               & $-1.872$\\
    \noalign{\smallskip}\cline{2-3}\noalign{\smallskip}
                    & $^{9}$Be$^+$              & $-1.873$\\
    \noalign{\smallskip}\cline{2-3}\noalign{\smallskip}
                    & $^{135}$Ba$^+$            & $-1.872$\\
    \noalign{\smallskip}\cline{2-3}\noalign{\smallskip}
                    & $^{137}$Ba$^+$            & $-1.872$\\
    \noalign{\smallskip}\cline{2-3}\noalign{\smallskip}
                    & $^{201}$Hg$^+$            & $-1.874$\\
\noalign{\smallskip}\hline\noalign{\smallskip}
    \multirow{4}{*}{$\frac{5}{2}$}  & $^{85}$Rb   & $-1.923$\\
    \noalign{\smallskip}\cline{2-3}\noalign{\smallskip}
                    & $^{25}$Mg$^+$             & $-1.923$\\
    \noalign{\smallskip}\cline{2-3}\noalign{\smallskip}
                    & $^{67}$Zn$^+$             & $-1.922$\\
    \noalign{\smallskip}\cline{2-3}\noalign{\smallskip}
                    & $^{173}$Yb$^+$            & $-1.922$\\
\noalign{\smallskip}\hline\noalign{\smallskip}
    \multirow{2}{*}{$\frac{7}{2}$}   & $^{133}$Cs & $-1.940$\\
    \noalign{\smallskip}\cline{2-3}\noalign{\smallskip}
                    & $^{43}$Ca$^+$             & $-1.939$\\
\noalign{\smallskip}\hline\noalign{\smallskip}
     $\frac{9}{2}$  &$^{87}$Sr$^+$              & $-1.947$\\
\hline
\end{tabular}
%
\end{table}
    \par
    The above-described procedure used for the evaluation does not include the influence of the fine structure. Because the BBRZ shift is a magnetic dipole interaction, the energy shift due to the hyperfine structure is the dominating term. We see these results using second-order perturbation theory,
\begin{equation}
  \Delta E\approx\sum_{e\neq g}\frac{|\langle e|\widehat{H}_{AF}|g\rangle|^2}{E(e)-E(g)},
\end{equation}
    where $\widehat{H}_{AF}$ is the atom--field interaction Hamiltonian, $e$ the excited state, and $g$ the ground state. The hyperfine splittings are usually several gigahertz, but the energy difference of the fine structure is of order 100~THz. In the present circumstance, the coupling of other fine-structure levels can be neglected.
    \par
    Itano and coworkers$^{[11]}$ analyzed the BBRZ shifts for any alkali atoms and alkali-like ions in the ground state, and presented a result of $-1.034\times10^{-17}[T(K)/300]^2$. Our results are fairly close to their result. In addition, we derived the BBRZ shifts for different atoms with different nuclear spins and different hyperfine splittings, and they may give some guidance in choosing atomic species for microwave atomic clocks.
    \par
    In Ref.~[12], the BBRZ shifts of cesium were calculated to be $-4.933\times10^{-15}[T(K)/300]^3$. Their results are quite different from Itano and coworkers and from our results. The key problem of Ref.~[12] may have been the failure to take into account the counter-rotating terms when using the RWA .
    \par
    In conclusion, this Letter gives a simple and detailed derivation of the BBRZ shifts for the ground states of alkali atoms and alkali-like ions through two-level system analysis. The calculation results consistent with previous works very well. The analytical results might have been very useful for experimental physicist doing microwave atomic clocks to estimate BBR Zeeman shifts and the further improvement of microwave atomic clocks.
\section{Acknowledgment}
\label{section:4}
    \par
    This project is supported by National Key Research and Development Program of China (No.2016YFA0302101), and the Initiative program of State Key Laboratory of Precision Measurement Technology and Instruments.\\
     \par
    We thank S.G.Porsev and Y.M.Yu for helpful discussions and suggestions. 
%
%
%
%

%
%
\end{document}